\documentclass[conference]{IEEEtran}

\pdfoutput=1

\newlength{\figwidth}
\usepackage[nosort]{cite}        
\usepackage{url} 
\usepackage[intlimits]{amsmath}
\usepackage{bbm}
\usepackage{graphicx}
\usepackage{paralist}
\usepackage[stretch=16,shrink=16,step=4]{microtype}
\usepackage{vmr-symbols-vecbold}
\usepackage{standard-macros}
\usepackage[caption=false,font=footnotesize]{subfig}

\usepackage{enumitem}

\makeatletter
\def\@IEEEinterspaceratioM{0.265}
\def\@IEEEinterspaceMINratioM{0.1651}
\def\@IEEEinterspaceMAXratioM{0.38}

\def\@IEEEinterspaceratioB{0.31}
\def\@IEEEinterspaceMINratioB{0.19}
\def\@IEEEinterspaceMAXratioB{0.38}
\@IEEEtunefonts
\makeatother
\safemath{\meno}{\beta}

\begin{document}

\IEEEoverridecommandlockouts

\title{One-Bit Massive MIMO: Channel Estimation \\ and High-Order Modulations}
\author{\IEEEauthorblockN{Sven Jacobsson\IEEEauthorrefmark{1}\IEEEauthorrefmark{2}, Giuseppe Durisi\IEEEauthorrefmark{1}, Mikael Coldrey\IEEEauthorrefmark{2}, Ulf Gustavsson\IEEEauthorrefmark{2}, Christoph Studer\IEEEauthorrefmark{3}}
\thanks{This work was partly supported by the Swedish Foundation for Strategic Research under grants SM13-0028 and ID14-0022, and by the Swedish Governmental Agency for Innovation Systems (VINNOVA) within the VINN Excellence center Chase.}
\IEEEauthorblockA{
\IEEEauthorrefmark{1}Chalmers University of Technology, Gothenburg, Sweden\\
\IEEEauthorrefmark{2}Ericsson Research, Sweden\\
\IEEEauthorrefmark{3}Cornell University, Ithaca, NY, USA
}}
%
%
\maketitle

\begin{abstract}
We investigate the information-theoretic throughout achievable on a fading communication link when the receiver is equipped with one-bit analog-to-digital converters (ADCs). 
The analysis is conducted for the setting where neither the transmitter  nor the receiver have \emph{a priori} information on the realization of the fading channels.
This means that channel-state information needs to be acquired at the receiver on the basis of the one-bit quantized channel outputs.
We show  that least-squares (LS) channel estimation combined with joint pilot and data processing is capacity achieving in the single-user, single-receive-antenna case.

We also investigate the achievable uplink throughput in a massive multiple-input multiple-output system where each element of the antenna array at the receiver base-station feeds a one-bit ADC.
We show that LS channel estimation and maximum-ratio combining are sufficient to support both multiuser operation and the use of high-order constellations. 
This holds in spite of the severe nonlinearity introduced by the one-bit ADCs.
\end{abstract}

\section{Introduction}
Digital signal processing (DSP) is an integral part of all modern communication systems. 
In order to process data digitally, the analog baseband signal has to be mapped to the digital domain. 
This requires conversion both in time (sampling) and amplitude (quantization).
The circuit that performs this last operation, known as analog-to-digital converter (ADC), is a necessary component in every system that includes DSP.
An ADC with frequency $f\sub{s}$ and resolution of $n$ bits maps the continuous-amplitude samples into a set of $ 2^n$ quantization levels, by operating $f\sub{s} 2^n$ conversion steps per second.  
A crucial problem with modern ADCs is that the power dissipated per conversion step (a.k.a. Walden's figure of merit~\cite{walden94-10a,walden99-04a}) increases dramatically for sampling rates higher than about $100\MHz$~\cite{murmann14-11a}. 
This implies that, for wideband communication systems, the resolution of the ADCs must be kept low to maintain a power budget that is within acceptable levels.

The one-bit resolution case, where the in-phase and the quadrature components of the continuous-valued received samples are quantized separately using one-bit ADCs (zero-threshold comparators),  is particularly attractive, because of the resulting low hardware complexity.
Indeed, in such a one-bit ADC architecture, there is no need for an automatic gain controller. 
Communication systems employing one-bit ADCs have been previously analyzed in the context of low-power ultra-wideband systems~\cite{hoyos05-07a,mezghani07-06a,landau14-09a}, and, more recently, in the context of millimeter-wave communication systems~\cite{mo14-10a,ulusoy13-01a}, and massive (or large-scale) multiple-input multiple-output (MIMO) systems~\cite{risi14-04a}.
In ultra-wideband and in millimeter-wave systems, the motivation for using one-bit ADCs is the large bandwidth of the transmitted signal. 
In massive MIMO systems, an additional reason is the massification of the number of radio-frequency chains at the base-station (BS), which makes the use of low-cost solutions---such as one-bit ADCs---attractive~\cite{risi14-04a}.

\paragraph*{Previous Results} 
\label{sec:previous_results}
A  receiver employing one-bit ADCs needs to cope with the severe nonlinearity introduced by such quantizers.
In their presence,  the signaling schemes and the receiver algorithms employed for the case of high-resolution quantizers become suboptimal.
The impact of the one-bit ADC nonlinearity on the performance of communication systems has been analyzed to some extent in the literature.
In~\cite{singh09-12a}, it is proven that 2-PAM is capacity achieving over a real-valued nonfading single-input single-output (SISO) Gaussian channel.
For  complex-valued Gaussian channels, QPSK turns out to be optimal.
For the general MIMO case, QPSK is not capacity achieving, and the capacity-achieving distribution is  unknown.
 
These results hold under the assumption that the one-bit ADC is a zero-threshold comparator. 
It turns out that in the low-SNR regime, a zero-threshold comparator is not optimal~\cite{koch13-09a}. 
The optimal strategy involves the use of \emph{flash signaling}~\cite[Def.~2]{verdu02-06a} and requires an optimization over the threshold value.  
Unfortunately, the power gain obtainable using this optimal strategy manifests itself only at extremely low values of spectral efficiency.
In the remainder of the paper, we therefore exclusively focus on the zero-threshold architecture.

Moving to Rayleigh-fading channels, QPSK is capacity achieving (again for the  SISO case) under the assumption that the receiver has somehow access to perfect channel state information (CSI)~\cite{krone10-08a}. 
  The assumption that perfect CSI is  available is, however, not realistic in the one-bit quantized case, since the nonlinear distortion caused by the one-bit quantizers makes it challenging to estimate the fading process perfectly.
  For the more practically relevant case when the channel is not known \emph{a priori} to the receiver, but must be learnt (for example, via pilot symbols), QPSK is optimal when the SNR exceeds a certain threshold that depends on the coherence time of the fading process~\cite{mezghani08-07a}. 
For SNR values that are below this threshold, on-off QPSK is capacity achieving~\cite{mezghani08-07a}.
  
In~\cite{mo14-10a}, crude high-SNR bounds are obtained for the capacity of one-bit-quantized MIMO fading channels, under the ideal assumption that perfect CSI is available to both the transmitter and the receiver.
Risi \emph{et al}.~\cite{risi14-04a} recently provided  a lower bound on the throughput achievable on  massive MIMO uplink channels, when the BS employs one-bit ADCs. 
The bound suggests that, in some scenarios, massive MIMO may  be robust against the coarse output quantization resulting from the use of one-bit ADCs.
However, as the lower bound obtained in~\cite{risi14-04a} is based on a suboptimal input distribution, i.e., QPSK, and a  suboptimal detection algorithm, i.e., least-squares (LS) channel estimation followed by maximal-ratio combining (MRC) or zero-forcing, its tightness is unclear.

All the results reviewed so far hold under the assumption of Nyquist-rate sampling at the receiver.
It is worth pointing out that 
Nyquist-rate sampling is not necessarily optimal in the presence of quantization at the receiver~\cite{shamai-shitz94-06a,koch10-11a}.
Indeed, higher information rates can be achieved by oversampling the received signal.
For example, for the complex AWGN case, high-order constellations such as 16-QAM can be supported in the SISO case if one allows for oversampling at the receiver~\cite{krone12-05a}.

\paragraph*{Contributions} 
\label{par:contributions}
Focusing on Nyquist-rate sampling, and on the scenario where neither the transmitter, nor the receiver have \emph{a priori} CSI, we investigate the rates achievable over Rayleigh block-fading MIMO channels when the receiver is equipped with one-bit ADCs.
Our contribution is twofold:
\begin{itemize}[leftmargin=*]
  \item For the SISO case,  we prove that LS channel estimation performed jointly on pilot and data symbols is capacity achieving. 
  In the infinite precision (no quantization) case, the benefit of \emph{joint pilot-data} (JPD) processing has been illustrated, e.g., in~\cite{coldrey08-01a,jindal09-06a,dorpinghaus12-05a}, where it is shown that joint processing yields a smaller gap to capacity compared to  separate pilot/data processing. 
  Our result shows that, in the one-bit ADC case, the gap to capacity is actually zero.
  Moreover,  LS estimation, although inferior to the optimal maximum a posteriori probability estimator (see~\cite{zymnis10-02a,studer14-09a}), suffices to achieve capacity when combined with JPD processing.
  
  \item We also consider the uplink of a massive MIMO system where single-antenna users communicate with a BS equipped with a large antenna array whose elements feed one-bit ADCs.
  Generalizing the analysis presented in~\cite{risi14-04a}, we show that MRC combined with LS channel estimation at the BS is sufficient to support both multi-user operations and the use of high-order constellation such as 16-QAM. 
  Furthermore, the rates achievable with 16-QAM turn out to exceed the ones reported in~\cite{risi14-04a} for QPSK, for SNR values as low as $-15$ dB, and for antenna arrays of $100$ elements or more.
  Our result suggests that temporal oversampling, as proposed in~\cite{krone12-05a}, can be replaced by spatial oversampling through the use of a massive antenna array at the BS.
\end{itemize}

\section{System Model}
\begin{figure}[t]
\centering
\includegraphics[width = 0.9\figwidth]{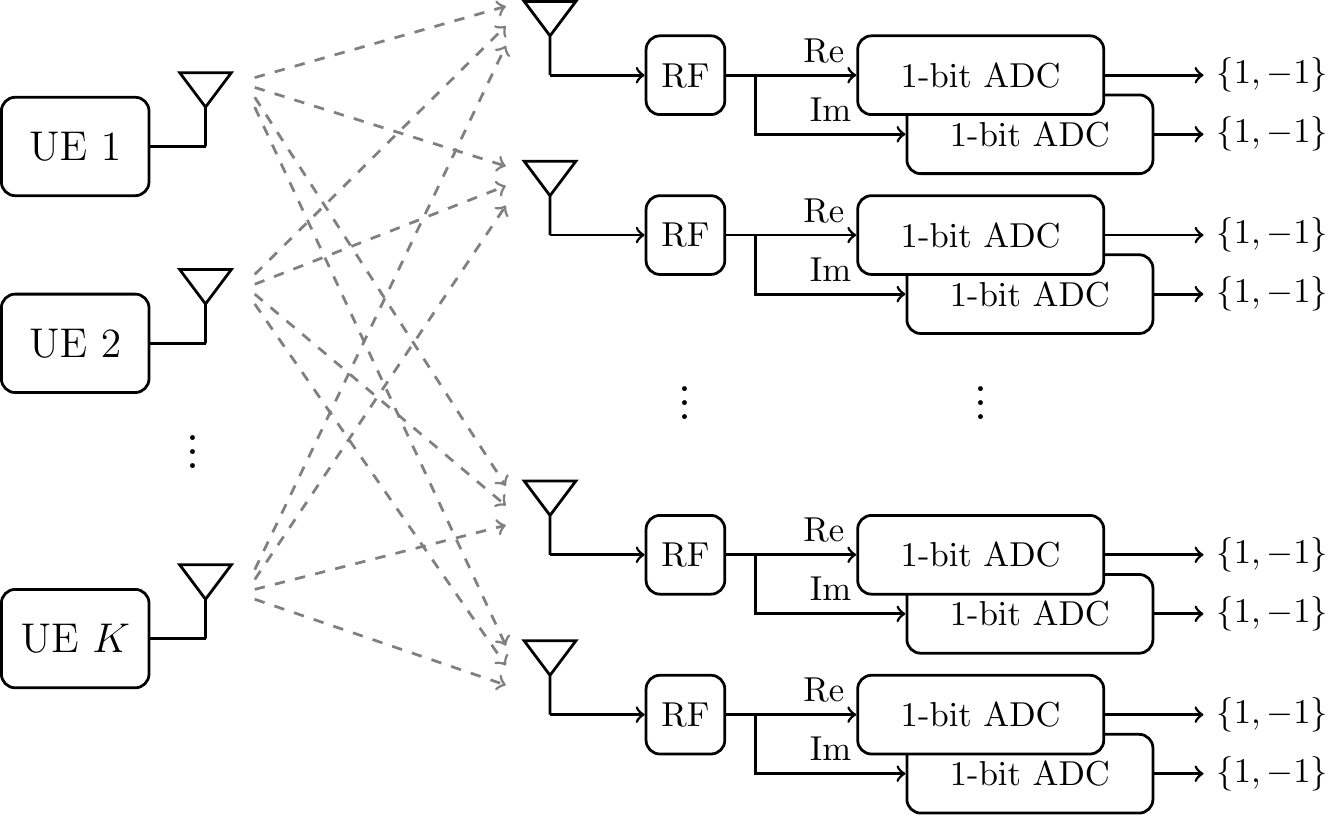}
\caption{One-bit massive MIMO uplink system model.}
\label{fig:system}
\end{figure}

We consider a single-cell uplink system as depicted Fig.~\ref{fig:system}, where $K$ single-antenna users are served by a BS that is equipped with an array of $N \gg K$ antennas. 
We model the subchannels between each transmit-receive antenna pair as a Rayleigh block-fading channel (see e.g.,~\cite{marzetta99-01a}), i.e., a channel that stays constant for~$T$ channel uses, and evolves independently across blocks of $T$ channel uses. 
We shall refer to $T$ as the channel coherence time (measured in channel uses). 
We further assume that the subchannels are mutually independent.
The discrete-time complex baseband received signal over all antennas within an arbitrary coherence block and before quantization, is modelled as
%
\begin{IEEEeqnarray}{rCL}
\rmatY &=& \matX \rmatH + \rmatW.
\end{IEEEeqnarray}
Here, $\matX \in \opC^{T\times K}$ denotes the channel input, $\rmatH \in \opC^{K \times N}$ is the channel matrix connecting the $K$ users to the $N$ BS antennas.
The entries of $\rmatH$ are independent and $\mathcal{CN}(0,1)$ distributed.
Furthermore, the matrix $\rmatW \in \opC^{T \times N}$, whose entries are  independent and $\mathcal{CN}(0,1)$ distributed,  stands for the AWGN.


The real and the imaginary components of the received signal at each antenna are quantized separately using a one-bit ADC.
Let $\setR = \{ 1+j, -1+j, -1-j,1-j\}$ be the set of possible quantization outcomes. 
It will be convenient to describe the joint operation of all $2N$ one-bit ADCs at the BS through the function $\setQ(\cdot): \opC^{T \times N} \rightarrow \setR^{T \times N}$
that maps the output matrix $\rmatY$ with entries $\{\rndy_{t,n}\}$ into the  quantized output matrix $\rmatR$ with entries $\{\rndr_{t,n}\}$ according to $\Re\{\rndr_{t,n}\} = \sign\{ \Re\{\rndy_{t,n}\} \}$ and $\Im\{\rndr_{t,n}\} = \sign\{ \Im\{\rndy_{t,n}\} \}$.
Using this convention, we can write the quantized output matrix as
\begin{IEEEeqnarray}{rCL}
\label{eq:systemMassiveMIMO}
\rmatR &=& \setQ \lefto(\rmatY \right) = \setQ \lefto(\matX \rmatH + \rmatW\right).
\end{IEEEeqnarray}

We shall consider the scenario where neither the users nor the BS are aware of the realizations of the channel matrix $\rmatH$ (no \emph{a priori} CSI), and where coding is performed over many coherence blocks.
For this scenario, the channel sum-rate capacity $C$ is
\begin{IEEEeqnarray}{rCL}\label{eq:sum-rate_capacity}
  C(\snr)=\sup I(\rmatX;\rmatR)
\end{IEEEeqnarray}
where the supremum is computed over all input probability distributions on $\rmatX$ for which the columns of $\rmatX$ are independent, and the following average-power constraint holds:
\begin{IEEEeqnarray}{rCL}
\Ex{}{\tr\{\rmatX\rmatX^H\}} & \le & KT\snr.
\end{IEEEeqnarray}
Since the noise variance is normalized to one, we can think of $\snr$ as the SNR.
The sum-rate capacity~\eqref{eq:sum-rate_capacity} is, in general, not known in closed form, even in the infinite-precision case, for which tight bounds have been recently reported in~\cite{devassy14-12a}.

\section{SISO Case}
\label{sec:SISO}
We focus in this section on the scenario where there is only a single active user in the system and where the BS has a single receive antenna.
For this case, the input-output relation~\eqref{eq:systemMassiveMIMO} reduces to
\begin{IEEEeqnarray}{rCL}\label{eq:io-siso}
\rvecr = \setQ(\rvecy) = \setQ(\vecx\rndh + \rvecw).
\end{IEEEeqnarray}
Here, $\vecx \in \opC^{T}$ and $\rvecw \in \opC^{T}$ are the input and noise vectors, respectively, and  $\rndh$ denotes the fading channel, which remains constant over the coherence block.
For this case, the capacity~\eqref{eq:sum-rate_capacity} is known \cite[Th.~1]{mezghani08-07a} and given by
\begin{equation}
\label{eq:CoptSISO}
C(\snr) = 
\begin{cases}
\dfrac{\snr}{\snr_c} R_{\text{QPSK}}(\snr_c), & \snr \le \rho_c \\
R_{\text{QPSK}}(\snr), & \rho > \snr_c.
\end{cases}
\end{equation}
Here, $R_{\text{QPSK}}$ denotes the rate achievable with QPSK,
\begin{IEEEeqnarray}{rCL}
\label{eq:RQPSK}
R_{\text{QPSK}}(\snr) 
&=& 2 + \frac{2}{T}\!\sum_{k=0}^T \! {T \choose k} \meno(k,T\!-k) \log_2 \! \meno(k,T\!-k) \IEEEeqnarraynumspace
\end{IEEEeqnarray}
where
\begin{equation}
\label{eq:gamma}
\meno(a,b) = \Ex{\rndg}{\Phi\left( -\rndg\sqrt{\snr} \right)^{a} \Phi\left( \rndg\sqrt{\snr} \right)^{b}}
\end{equation}
with $\rndg\distas\mathcal{N}(0,1)$ and $\Phi(x)$ denoting the cumulative distribution function of a standard normal random variable.
Furthermore, the SNR threshold $\snr_c$ in~\eqref{eq:CoptSISO} is the solution of the following optimization problem: 
\begin{IEEEeqnarray}{rCL}
\label{eq:SNRoptSISO}
\snr_c = \argmax_{\snr \ge 0} \dfrac{1}{\snr}R_{\text{QPSK}}(\snr).
\end{IEEEeqnarray}

A common approach to transmitting information over fading channels whose realizations are not known \emph{a priori} to the receiver is to reserve a certain number of channel uses at the beginning of each coherence block for the transmission of pilot symbols~\cite{tong04-11a}.
These pilot symbols are then used at the receiver to estimate the fading channel.
Assume that $0\leq P\leq T$ pilots are used and let $\vecx^{(p)}$ be the $P$-dimensional vector containing these pilot symbols. 
Similarly, let $\rvecr^{(p)}$ be the corresponding one-bit quantized channel output.
The pair $(\vecx^{(p)},\rvecr^{(p)})$ can be used at the receiver to estimate the channel $\rndh$. 
As in~\cite{risi14-04a}, we shall focus on LS estimation because of its low complexity.
When the pilots are QPSK symbols, the LS estimate~$\hat{\rndh}$ of~$\rndh$ is 
\begin{IEEEeqnarray}{rCL}\label{eq:ls-estimate-siso}
  \hat{\rndh} = \frac{1}{P \sqrt{\snr}} \herm{(\vecx^{(p)})}\vecr^{(p)}.
\end{IEEEeqnarray}
Under the assumption that the $T-P$  data symbols are drawn independently from the same input distribution, one obtains then the following lower bound on $C$:
\begin{IEEEeqnarray}{rCL}
  \IEEEeqnarraymulticol{3}{l}{C(\snr)\geq \frac{T-P}{T}I(\rndx;\rndr \given \hat{\rndh})}\label{eq:pilot-based_bound_step_1}\\
  %
  \!&=& 2\frac{T\!-\!P}{T}\!\lefto[1\!-\!\sum_{\ell=0}^P \! {P\choose \ell} \meno(\ell,P\!-\!\ell) H_b\!\lefto(\frac{\meno(\!\ell\!+\!1,P\!-\!\ell)}{\meno(\ell,P\!-\!\ell)}\!\right)\!\right]\!\!. \IEEEeqnarraynumspace
  \label{eq:pilot-based_bound}
\end{IEEEeqnarray}
Here, $H_b(\cdot)$ denotes the binary entropy function. The inequality in~\eqref{eq:pilot-based_bound_step_1} follows from standard manipulations on the mutual information (see, e.g.,~\cite{dorpinghaus12-05a}); the equality~\eqref{eq:pilot-based_bound} is proven in Appendix~\ref{app:proof_of_LS}.

In Fig.~\ref{fig:rateLS}, we plot the capacity~\eqref{eq:CoptSISO} and the pilot-based LS-estimation lower bound~\eqref{eq:pilot-based_bound} for the case $\snr=10$ dB.
Note that,  for this $\snr$ value, $C(\snr)=R\sub{QPSK}(\snr)$ for all $T$.
The number of pilots in \eqref{eq:pilot-based_bound} is optimized for each value of $T$.
For reference, we also depict the perfect receiver-CSI capacity~\cite{krone10-08a}.
As $T$ grows large, the gap between~\eqref{eq:CoptSISO} and the perfect receiver-CSI capacity decreases.
The gap between the pilot-based LS-estimation lower bound~\eqref{eq:pilot-based_bound} and capacity~\eqref{eq:CoptSISO} is essentially constant over the considered range of $T$ values.
One exception is the case $T=2$, for which there seems to be no gap. 
Indeed, the following result holds (see \cite[Lem.~1]{jacobsson15-01a}).
\begin{lem}
\label{lemma}
The RHS of~\eqref{eq:pilot-based_bound} coincides with the rate acheivable with QPSK~\eqref{eq:RQPSK} for the case $T=2$.
\end{lem}
\begin{figure}[t]
\centering
\includegraphics[width = .997\figwidth]{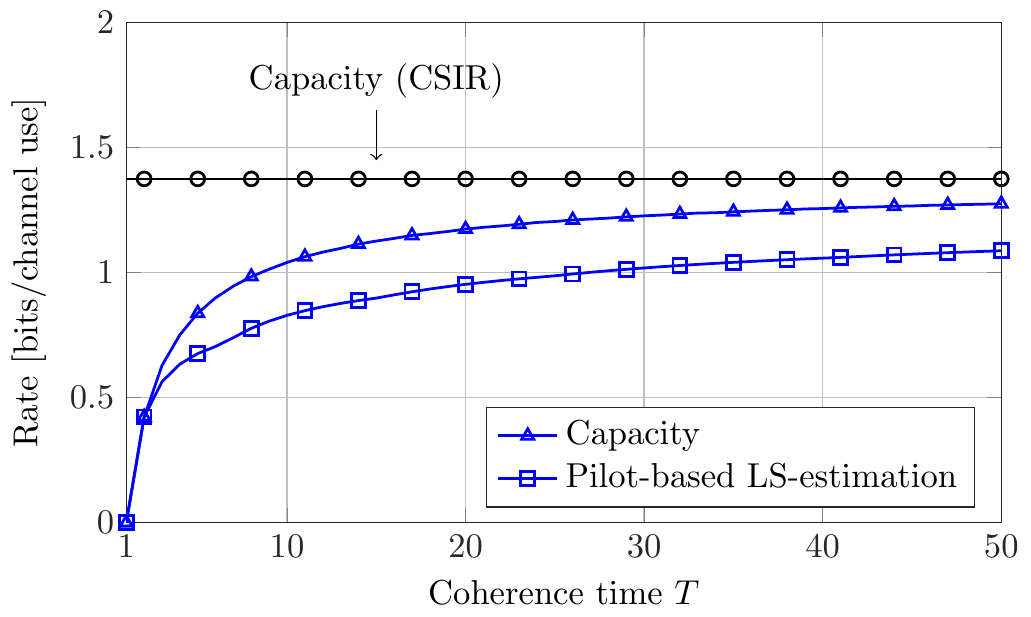}
\caption{Comparison of capacity~\eqref{eq:CoptSISO} and the pilot-based LS-estimation lower bound~\eqref{eq:pilot-based_bound} for the SISO case when $\snr = 10$ dB.}
\label{fig:rateLS}
\end{figure}

It is well known that the pilot-based lower bound~\eqref{eq:pilot-based_bound} can be improved by using also the channel outputs corresponding to the data symbols to improve the channel estimate~\cite{coldrey08-01a,jindal09-06a,dorpinghaus12-05a}. 
This approach is sometimes referred to as JPD processing.
Perhaps surprisingly, for the one-bit quantization case, LS estimation combined with JPD processing turns out to be optimal, as formalized in the following theorem.
\begin{thm}\label{thm:data-aided-LS}
For the channel~\eqref{eq:io-siso}, LS estimation combined with JPD processing achieves the channel capacity~\eqref{eq:CoptSISO}.
\end{thm}
\begin{IEEEproof}
  See Appendix~\ref{app:proof_of_theorem}.
\end{IEEEproof}

This result implies that if one allows for JPD processing, LS channel estimation is optimal, and there is no need to use more sophisticated channel-estimation techniques such as the one recently proposed in \cite{Wen15-01a}.

\section{Massive MIMO Case}

Motivated by the results obtained for the SISO case, we now assess the rates achievable with LS estimation in a multiuser massive MIMO uplink scenario.
To limit the receiver complexity, we shall only consider the pilot-based version of the LS estimation algorithm (no JPD processing).
Indeed, JPD processing is in general computationally demanding~\cite{dorpinghaus12-05a} and may be not suitable for massive MIMO. We shall also assume that the receiver employs MRC to separate the information streams associated with the different users. 

\begin{figure}[t]
\centering
\subfloat[$N=40$ antennas, $\snr = 0$ dB.]{
\includegraphics[width = .475\figwidth]{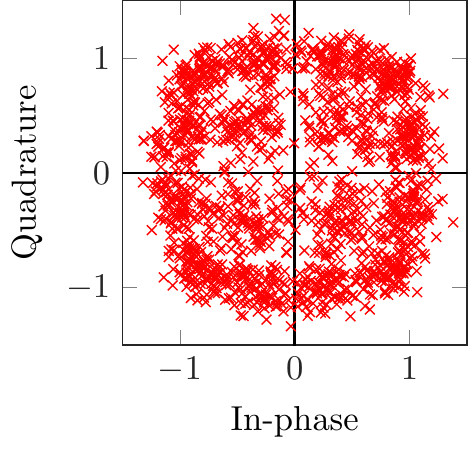}
\label{fig:const40_0dB}
}
\subfloat[$N=400$ antennas, $\snr = 0$ dB.]{
\includegraphics[width = .475\figwidth]{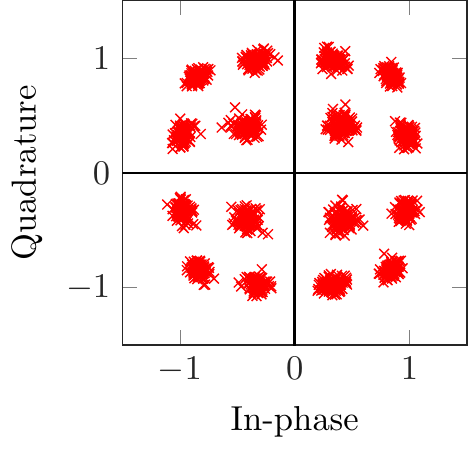}
\label{fig:const400_0dB}
}
\\
\subfloat[$N=400$ antennas, $\snr = 20$ dB.]{
\includegraphics[width = .475\figwidth]{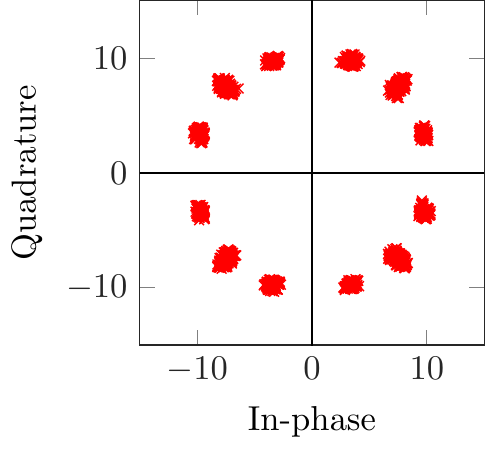}
\label{fig:const400_20dB}
}
\subfloat[$N=400$ antennas, $\snr = 20$ dB, fully correlated channel coefficients.]{
\includegraphics[width = .475\figwidth]{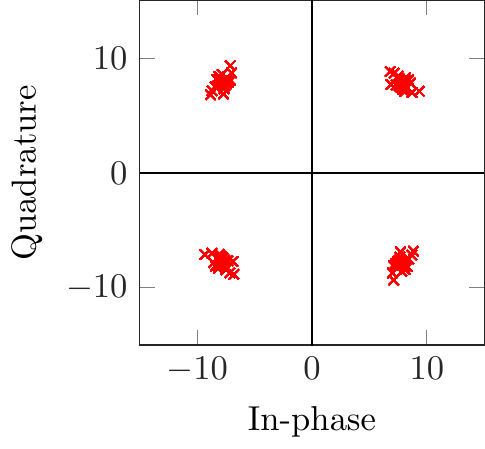}
\label{fig:const400_20dB_correlated}
}
\caption{Single-user MRC outputs with LS channel estimation as a function of the number of receive antennas and the SNR; 16-QAM inputs.}
\label{fig:const}
\end{figure}

We assume that the users coordinate the pilot-transmission phase. 
Specifically, they transmit their pilot sequences in a round-robin fashion.
The channel estimates are then used to obtain the MRC coefficients.
%
%
Differently from~\cite{risi14-04a}, where a similar setup is considered, we focus on high-order modulations and not only on QPSK.
Indeed, although QPSK is optimal in the SISO case, the use of multiple antennas at the receiver opens up the possibility to use high-order modulation formats.
This is demonstrated in Fig.~\ref{fig:const} where we plot the MRC receiver output corresponding to 16-QAM data symbols for the case when a single user, alone in the cell, transmits also $P = 20$ pilots to let the BS acquire LS channel estimates.
As the size of the receiver antenna array increases, the 16-QAM constellation becomes progressively distinguishable (see Fig.~\ref{fig:const400_0dB}), provided that $\snr$ is not too high.

Additive noise is one of the factors that enables the detection of the 16-QAM constellation; the other is the independent phase of the fading coefficients associated with each receive antenna. Recall in fact that, due to the one-bit quantizer, the quantized output at each receive antenna belongs to the set $\setR$ of cardinality $4$.
These 4 possible channel outputs are then averaged by the MRC filter to produce a channel output (a scalar) that belongs to an alphabet with much higher cardinality.
The cardinality depends on the number of pilots and receive antennas. 
The key observation is that the inner points of the 16-QAM constellation are more likely to be erroneously detected.
This results in a smaller averaged value after MRC than for the outer constellation points.

To highlight the importance of the additive noise, we consider in Fig.~\ref{fig:const400_20dB} the case $\snr=20$ dB.
Since the additive noise is negligible, all 16-QAM constellation points are detected correctly with high probability. 
As a result, the output of the MRC filter lies approximately on a circle, which suggests that the amplitude of the transmitted signal cannot be used to convey information.
When the noise is negligible and all fading coefficients are fully correlated, the constellation collapses to a noisy QPSK diagram (Fig.~\ref{fig:const400_20dB_correlated}). 
In this case, high-order modulations are not supported by the channel.

The achievable rate $R^{(k)}_{\text{MRC}}$ for user $k = 1, \dots, K$ with LS estimation and MRC is
\begin{equation}
\label{eq:achRate_MassiveMIMO}
R^{(k)}_{\text{MRC}}(\snr) = \frac{T-P}{T}I(\rndx^{(k)};\tilde{\rndx}^{(k)} \given \hat{\matH}).
\end{equation}
The mutual information between the channel input $\rndx^{(k)}$ and the MRC receiver output $\tilde{\rndx}^{(k)}$ can be computed by mapping $\tilde{\rndx}^{(k)}$ to points over a regular grid in the complex plane as described in~\cite{risi14-04a}.
With this technique, one obtains a lower bound on $R^{(k)}_{\text{MRC}}(\snr)$~\cite[p.~3503]{arnold06-08a} that becomes increasingly tight as the grid spacing is driven to zero.
The conditional probability mass functions needed for the evaluation of the mutual information are computed using Monte-Carlo techniques.\footnote{The numerical routines that are used to evaluate~\eqref{eq:achRate_MassiveMIMO} can be downloaded at \url{https://github.com/infotheorychalmers/one-bit_massive_MIMO}.}
Since all the users in the system are assumed to be statistically equivalent, we have that
\begin{IEEEeqnarray}{rCL}
  C(\snr)\geq K R^{(1)}_{\text{MRC}}(\snr).
\end{IEEEeqnarray}

\begin{figure}[t]
\centering
\includegraphics[width = .997\figwidth]{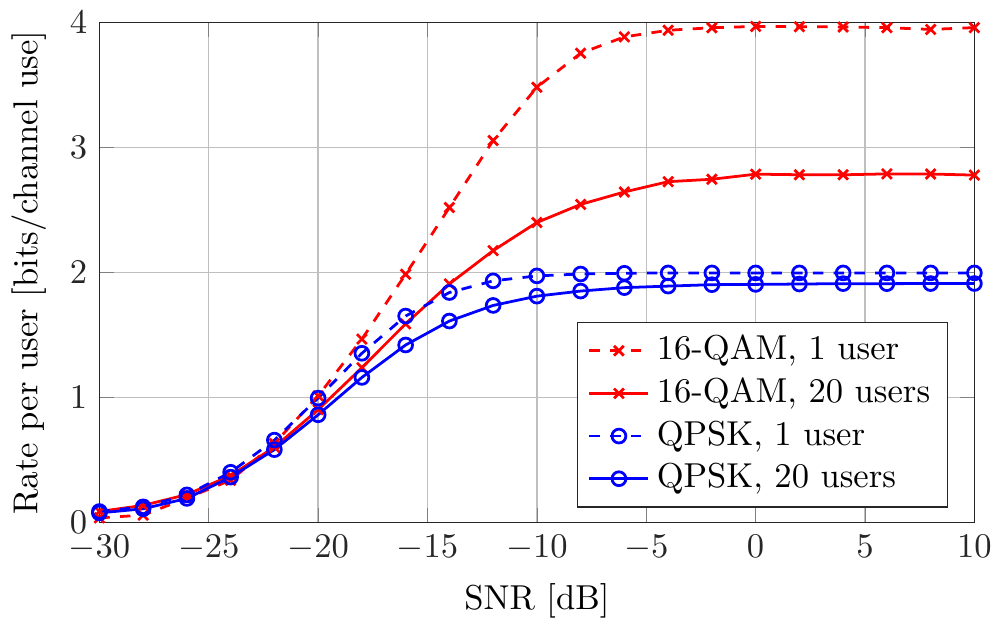}
\caption{Per-user achievable rate with LS estimation and MRC as a function of $\snr$; $T = 1000$, $N = 400$; the number of pilots $P$ is optimized for each value of $\snr$.}
\label{fig:rate_vs_SNR}
\end{figure}

In Fig.~\ref{fig:rate_vs_SNR} we compare the rates achievable with QPSK and 16-QAM as a function of $\snr$.
The number of receive antennas is $N=400$, the coherence time is  $T=1000$, and we consider both the case when the number of users $K$ is $1$ and $20$.
The number of transmitted pilots is  optimized for every $\snr$ value.
We see that 16-QAM outperforms QPSK already at SNR values as low as $-15$ dB. 
Furthermore, the full  16-QAM rate of 4 bits per channel use can be achieved in the single-user case for large SNR values. 
Note that if $\snr$ is further increased, the 16-QAM rate starts decreasing, because the constellation collapses to a circle (cf. Fig.~\ref{fig:const400_20dB}).
Note also that, when QPSK is used, the difference in achievable rate between the case $K=1$ and $K=20$ is marginal---an observation already reported in~\cite{risi14-04a}.
On the contrary, the difference in achievable rates between single- and multi-user case is more pronounced when 16-QAM is used.
This suggests that, with 16-QAM, the system becomes interference limited, and that the one-bit quantizers partly destroy the orthogonality between the fading channels associated with different users.

\begin{figure}[t]
\centering
\includegraphics[width = .997\figwidth]{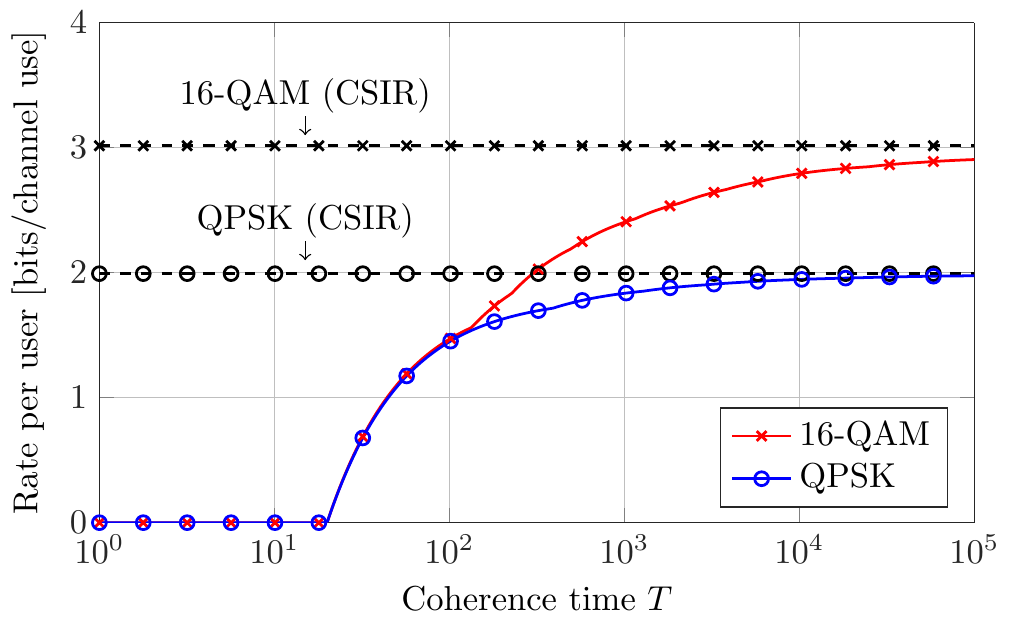}
\caption{Per-user achievable rate with LS estimation and MRC per user as a function of $T$; $N = 400$, $\snr = -10$ dB, $K = 20$; the number of pilots $P$ is optimized for each value of $T$.}
\label{fig:rate_vs_T}
\end{figure}
\begin{figure}[t]
\centering
\includegraphics[width = .997\figwidth]{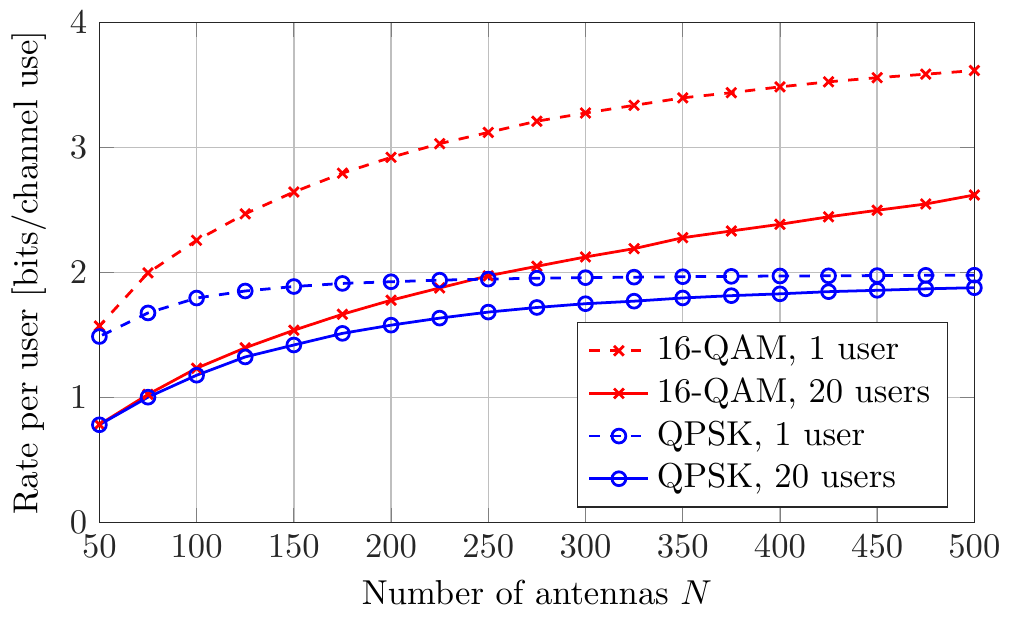}
\caption{Per-user achievable rates as a function of the number of antennas; $T = 1000$, $\snr = -10$ dB; the number of pilots $P$ is optimized for each value of $N$.}
\label{fig:rate_vs_antennas}
\end{figure}

In Fig.~\ref{fig:rate_vs_T}, we plot the per-user achievable rates  as a function of~$T$ for  $\snr = -10$ dB,  $N = 400$, and $K = 20$. 
The number of pilot symbols is again optimized for each value of $T$. 
We also depict the achievable rates for the perfect receiver-CSI case. 
Similarly to the SISO case, as $T$ increases, the per-user achievable rates approach the perfect-receive CSI rate. 
However, this convergence occurs at a much slower pace than for the infinite-precision case (cf.~\cite{rusek12-01a,devassy14-12a}).
This suggests that the one-bit ADC architecture may be unsuitable for high-mobility scenarios.
Note also that the achievable rate is zero when $T \le 20$. 
In fact, when orthogonal pilot sequences are transmitted, at least $20$ pilot symbols are required when $K=20$.
%


Finally, in Fig.~\ref{fig:rate_vs_antennas} we plot the per-user achievable rates as a function of the number of antennas. Here, $\snr = -10$ dB, and $T=1000$. 
As in the previous cases, the number of pilot symbols is optimized for each value of $N$. 
We note that 16-QAM  outperforms QPSK also when the number of receive antennas is much smaller than $400$.
%
We note also that, when QPSK is used, the achievable rate saturate rapidly as the number of receive antennas is increased.

\section{Conclusions}

We have analyzed the performance of a one-bit quantized receiver architecture operating over a Rayleigh block-fading channel whose realizations are not known \emph{a priori} to transmitter and receiver.
We have demonstrated that, for the SISO case, a signaling scheme based on LS estimation and JPD processing is capacity achieving. 
For the one-bit massive MIMO case, we have shown that, in contrast to the SISO case, high-order constellations can be used to convey information at higher rates than with QPSK.
This holds in spite of the nonlinearity introduced by the one-bit quantizer and in spite of the multiuser interference.
Similar results hold for the case when zero-forcing instead of MRC is used (see~\cite{jacobsson15-01a}).
Note also that constellations that are optimized for the nonlinearity introduced by the one-bit quantizers may yield higher achievable rates than the 16-QAM constellation analyzed in this paper.
Extension of our analysis to the case when the ADCs have 2 or 3 bits of resolution is also of interest.

\appendices
\section{Proof of~(12)}\label{app:proof_of_LS}
Both the pilot and the data symbols are assumed to belong to a QPSK constellation.
By symmetry, the rates achievable on the SISO channel~\eqref{eq:io-siso} with QPSK inputs are twice as high as the rates achievable with BPSK.
Hence, in the remainder of the proof, we shall consider a real-valued version of the channel input-output relation~\eqref{eq:io-siso}, where $\rndh$ and $\rvecw$ are real Gaussian, and the input vector $\rvecx$ consists of BPSK symbols.
Let $\ell$ denote the number of sign mismatches between the BPSK vectors $\vecx^{(p)}$ and the quantized received vector $\vecr^{(p)}$.
Since for the real case there exists a one-to-one relation between the LS estimate $\hat{\rndh}$ in~\eqref{eq:ls-estimate-siso} and $\ell$, we conclude that $I(\rndx;\rndr\given \hat{\rndh})=I(\rndx;\rndr\given \ell)$.
To evaluate this mutual information, we need the conditional probability mass function $p_{\rndr | \rndx,\rndell}$, which can be expressed as follows
\begin{IEEEeqnarray}{rCL}\label{eq:conditional_r}
  p_{\rndr | \rndx,\rndell}(r | x, \ell)=\frac{\Ex{\rndh}{p_{\rndell\given h}(\ell\given h)\Phi(r h x)}}{p_\rndell(\ell)}.
\end{IEEEeqnarray}
Using a similar approach as the one detailed in~\cite{mezghani08-07a}, one can show that
\begin{IEEEeqnarray}{rCL}\label{eq:conditional_l}
  p_{\rndell\given \rndh}(\ell\given h)= {P \choose \ell} \Phi(-h\sqrt{\snr})^{\ell} \Phi(h\sqrt{\snr})^{P-\ell}
\end{IEEEeqnarray}
and
\begin{IEEEeqnarray}{rCL}\label{eq:marginal_l}
  p_{\rndl}(\ell)= {P \choose \ell} \meno(\ell,P-\ell).
\end{IEEEeqnarray}
Substituting~\eqref{eq:conditional_l} and~\eqref{eq:marginal_l} into~\eqref{eq:conditional_r}, and then using the definition of mutual information, and that 
\begin{IEEEeqnarray}{rCL}\label{eq:properties_meno_coeff}
  \meno(\ell+1,P-l)+\meno(\ell,P-\ell+1)=\meno(\ell,P-\ell)
\end{IEEEeqnarray}
one obtains~\eqref{eq:pilot-based_bound} (see~\cite[App.~B]{jacobsson15-01a} for details).

\section{Proof of Theorem~2}\label{app:proof_of_theorem}
Our JPD processing lower bound is based on the following scheme. The first transmit symbol in each coherence block is a pilot symbol. 
To decode the $n$th symbol in the block, we rely on the LS channel estimate obtained on the basis of the past $n-1$ symbols (1 pilot symbol and $n-2$ data symbols).
This scheme yields the following achievable rate:
\begin{IEEEeqnarray}{rCL}\label{eq:rates_joint_pilot_data}
  R^{(T)}\sub{JPD}(\snr)=\frac{1}{T}\sum_{n=2}^{T} I\bigl(\rndx_n;\rndr_n\given \hat{\rndh}(n-1)\bigr).
\end{IEEEeqnarray}
Here, 
we have indicated the channel estimate by $\hat{h}(n-1)$ to clarify the number of input symbols that are used to estimate the channel. 
As in Appendix~\ref{app:proof_of_LS}, it is sufficient to focus on a real-valued version of the channel input-output relation~\eqref{eq:io-siso}, where $\rndh$ and $\rvecw$ are real Gaussian, and the input vector $\rvecx$ consists of BPSK symbols.

To establish Theorem~\ref{thm:data-aided-LS}, it is then sufficient to show that~\eqref{eq:rates_joint_pilot_data} coincides with
\begin{IEEEeqnarray}{rCL}
\label{eq:rates_BPSK}
R^{(T)}\sub{BPSK}(\snr)\!&=&\! 1 \!+\! \frac{1}{T} \! \sum_{k=0}^T \! {T \choose k} \meno(k,T\!-k) \log_2 \! \meno(k,T\!-k). \IEEEeqnarraynumspace
\end{IEEEeqnarray}
The final result~\eqref{eq:CoptSISO} is then established by replacing BPSK with on-off BPSK. See~\cite{mezghani08-07a} for the details.
Our proof that~\eqref{eq:rates_joint_pilot_data} coincides with~\eqref{eq:rates_BPSK} is by induction.
We start by noting that when $T=2$,~\eqref{eq:rates_joint_pilot_data} coincides with the RHS of~\eqref{eq:pilot-based_bound}.
Equality between~\eqref{eq:rates_joint_pilot_data} and~\eqref{eq:rates_BPSK} for this case then follows from Lemma~\ref{lemma}.

We now assume that JPD processing achieves~\eqref{eq:rates_BPSK} for a given coherence time $T$. 
We need to prove that the same holds when the coherence time is $T+1$.
Note that 
\begin{IEEEeqnarray}{rCL}
  R^{(T+1)}\sub{JPD}\!(\snr)\!&=&\!\frac{1}{T\!+\!1}\!\Bigl[ T R^{(T)}\sub{JPD}(\snr) 
  + I\bigl(x_{T+1};r_{\footnotesize{T+1}} \given \hat{h}(T)\bigr)\Bigr]. \IEEEeqnarraynumspace
  \label{eq:induction_step}
\end{IEEEeqnarray}
%
%
By the induction hypothesis, we can replace $R^{(T)}\sub{JPD}(\snr)$ by $R^{(T)}\sub{BPSK}(\snr)$.
Furthermore, we can replace the mutual information on the RHS of~\eqref{eq:induction_step} with the RHS of~\eqref{eq:pilot-based_bound}, evaluated for $P=T$.
The desired result then follows by using~\eqref{eq:properties_meno_coeff}, the following binomial equality
\begin{IEEEeqnarray}{rCL}
  {n \choose k}= {n-1 \choose k-1} + {n-1 \choose k}
\end{IEEEeqnarray}
and by performing simple algebraic manipulations.


\bibliographystyle{IEEEtran}
\bibliography{IEEEabrv,publishers,confs-jrnls,giubib}

\begin{thebibliography}{10}
\providecommand{\url}[1]{#1}
\csname url@samestyle\endcsname
\providecommand{\newblock}{\relax}
\providecommand{\bibinfo}[2]{#2}
\providecommand{\BIBentrySTDinterwordspacing}{\spaceskip=0pt\relax}
\providecommand{\BIBentryALTinterwordstretchfactor}{4}
\providecommand{\BIBentryALTinterwordspacing}{\spaceskip=\fontdimen2\font plus
\BIBentryALTinterwordstretchfactor\fontdimen3\font minus
  \fontdimen4\font\relax}
\providecommand{\BIBforeignlanguage}[2]{{%
\expandafter\ifx\csname l@#1\endcsname\relax
\typeout{** WARNING: IEEEtran.bst: No hyphenation pattern has been}%
\typeout{** loaded for the language `#1'. Using the pattern for}%
\typeout{** the default language instead.}%
\else
\language=\csname l@#1\endcsname
\fi
#2}}
\providecommand{\BIBdecl}{\relax}
\BIBdecl

\bibitem{walden94-10a}
R.~H. Walden, ``{Analog-to-digital converter technology comparison},'' in
  \emph{IEEE GaAs IC Symposium Technical Digest}, Philadelphia, PA, Oct. 1994,
  pp. 217--219.

\bibitem{walden99-04a}
------, ``{Analog-to-digital converter survey and analysis},'' \emph{{IEEE} J.
  Sel. Areas Commun.}, vol.~17, no.~4, pp. 539--550, Apr. 1999.

\bibitem{murmann14-11a}
\BIBentryALTinterwordspacing
B.~Murmann, ``{ADC Performance Survey 1997-2014}.'' [Online]. Available:
  \url{http://web.stanford.edu/~murmann/adcsurvey.html}
\BIBentrySTDinterwordspacing

\bibitem{hoyos05-07a}
S.~Hoyos, B.~Sadler, and G.~Arce, ``Monobit digital receivers for ultrawideband
  communications,'' \emph{{IEEE} Trans. Wireless Commun.}, vol.~4, no.~4, pp.
  1337--1344, Jul. 2005.

\bibitem{mezghani07-06a}
A.~Mezghani and J.~Nossek, ``On ultra-wideband {MIMO} systems with 1-bit
  quantized outputs: Performance analysis and input optimization,'' in
  \emph{Proc. IEEE Int. Symp. Inf. Theory (ISIT)}, Nice, France, Jun. 2007, pp.
  1286--1289.

\bibitem{landau14-09a}
L.~Landau and G.~Fettweis, ``On reconstructable ask-sequences for receivers
  employing 1-bit quantization and oversampling,'' in \emph{{IEEE} Int. Conf.
  Ultra-WideBand (ICUWB)}, Paris, France, Sep. 2014, pp. 180--184.

\bibitem{mo14-10a}
\BIBentryALTinterwordspacing
J.~Mo and R.~W. {Heath Jr}, ``Capacity analysis of one-bit quantized {MIMO}
  systems with transmitter channel state information,'' Oct. 2014. [Online].
  Available: \url{http://arxiv.org/abs/1410.7353}
\BIBentrySTDinterwordspacing

\bibitem{ulusoy13-01a}
A.~{Ulusoy et al.}, ``A 60 {GHz} multi-{Gb/s} system demonstrator utilizing
  analog synchronization and 1-bit data conversion,'' in \emph{IEEE Topical
  Meeting on Silicon Monolithic Integrated Circuits in RF Systems (SiRF)},
  Austin, TX, Jan. 2013, pp. 99--101.

\bibitem{risi14-04a}
\BIBentryALTinterwordspacing
C.~Risi, D.~Persson, and E.~G. Larsson, ``Massive {MIMO} with 1-bit {ADC},''
  Apr. 2014. [Online]. Available: \url{http://arxiv.org/abs/1404.7736}
\BIBentrySTDinterwordspacing

\bibitem{singh09-12a}
J.~Singh, O.~Dabeer, and U.~Madhow, ``On the limits of communication with
  low-precision analog-to-digital conversion at the receiver,'' \emph{{IEEE}
  Trans. Commun.}, vol.~57, no.~12, pp. 3629--3639, Dec. 2009.

\bibitem{koch13-09a}
T.~Koch and A.~Lapidoth, ``At low {SNR}, asymmetric quantizers are better,''
  \emph{{IEEE} Trans. Inf. Theory}, vol.~59, no.~9, pp. 5421--5445, Sep. 2013.

\bibitem{verdu02-06a}
S.~Verd{\'u}, ``Spectral efficiency in the wideband regime,'' \emph{{IEEE}
  Trans. Inf. Theory}, vol.~48, no.~6, pp. 1319--1343, Jun. 2002.

\bibitem{krone10-08a}
S.~Krone and G.~Fettweis, ``Fading channels with 1-bit output quantization:
  Optimal modulation, ergodic capacity and outage probability,'' in \emph{Proc.
  IEEE Inf. Theory Workshop (ITW)}, Dublin, Ireland, Aug. 2010.

\bibitem{mezghani08-07a}
A.~Mezghani and J.~Nossek, ``Analysis of {Rayleigh}-fading channels with 1-bit
  quantized output,'' in \emph{Proc. IEEE Int. Symp. Inf. Theory (ISIT)},
  Toronto, ON, Canada, Jul. 2008, pp. 260--264.

\bibitem{shamai-shitz94-06a}
S.~Shamai~(Shitz), ``Information rates by oversampling the sign of a
  bandlimited process,'' \emph{{IEEE} Trans. Inf. Theory}, vol.~40, no.~4, pp.
  1230--1236, Jun. 1994.

\bibitem{koch10-11a}
T.~Koch and A.~Lapidoth, ``Increased capacity per unit-cost by oversampling,''
  in \emph{IEEE 26th Convention of Electrical and Electronics Engineers in
  Israel (IEEEI),}, Eliat, Israel, Nov. 2010, pp. 684 --688.

\bibitem{krone12-05a}
S.~Krone and G.~Fettweis, ``Capacity of communications channels with 1-bit
  quantization and oversampling at the receiver,'' in \emph{{IEEE} Sarnoff
  Symp. (SARNOFF)}, Newark, NJ, May 2012.

\bibitem{coldrey08-01a}
M.~Coldrey and P.~Bohlin, ``Training-based mimo systems: Part ii--improvements
  using detected symbol information,'' \emph{{IEEE} Trans. Signal Process.},
  vol.~56, no.~1, pp. 296--303, Jan. 2008.

\bibitem{jindal09-06a}
N.~Jindal, A.~Lozano, and T.~Marzetta, ``What is the value of joint processing
  of pilots and data in block-fading channels?'' in \emph{Proc. IEEE Int. Symp.
  Inf. Theory (ISIT)}, Seoul, Korea, Jun. 2009, pp. 2189--2193.

\bibitem{dorpinghaus12-05a}
M.~Dorpinghaus, A.~Ispas, and H.~Meyr, ``On the gain of joint processing of
  pilot and data symbols in stationary rayleigh fading channels,'' \emph{{IEEE}
  Trans. Inf. Theory}, vol.~58, no.~5, pp. 2963--2982, May 2012.

\bibitem{zymnis10-02a}
A.~Zymnis, S.~Boyd, and E.~Cand\'es, ``Compressed sensing with quantized
  measurements,'' \emph{{IEEE} Signal Process. Lett.}, vol.~17, no.~2, pp.
  149--152, Feb. 2010.

\bibitem{studer14-09a}
C.~Studer and G.~Durisi, ``Quantized massive {MIMO}-{OFDM} uplink,'' 2015, in
  preparation.

\bibitem{marzetta99-01a}
T.~L. Marzetta and B.~M. Hochwald, ``Capacity of a mobile multiple-antenna
  communication link in {Rayleigh} flat fading,'' \emph{{IEEE} Trans. Inf.
  Theory}, vol.~45, no.~1, pp. 139--157, Jan. 1999.

\bibitem{devassy14-12a}
\BIBentryALTinterwordspacing
R.~Devassy, G.~Durisi, J.~\"{O}stman, W.~Yang, T.~Eftimov, and Z.~Utkovski,
  ``Finite-{SNR} bounds on the sum-rate capacity of {Rayleigh} block-fading
  multiple-access channels with no a priori {CSI},'' Dec. 2014. [Online].
  Available: \url{http://arxiv.org/abs/1501.01957}
\BIBentrySTDinterwordspacing

\bibitem{tong04-11a}
L.~Tong, B.~M. Sadler, and M.~Dong, ``Pilot-assisted wireless transmissions,''
  \emph{{IEEE} Signal Process. Mag.}, vol.~21, no.~6, pp. 12--25, Nov. 2004.

\bibitem{jacobsson15-01a}
S.~Jacobsson, ``Throughput analysis of massive {MIMO} uplink with one-bit
  {ADCs},'' Master's thesis, Chalmers University of Technology, Gothenburg,
  Sweden, Jan. 2015.

\bibitem{Wen15-01a}
\BIBentryALTinterwordspacing
C.-K. Wen, S.~Jin, K.-K. Wong, C.-J. Wang, and G.~Wu, ``Joint channel-and-data
  estimation for large-{MIMO} systems with low-precision {ADCs},'' Jan. 2015.
  [Online]. Available: \url{http://arxiv.org/abs/1501.05580}
\BIBentrySTDinterwordspacing

\bibitem{arnold06-08a}
D.~Arnold, H.-A. Loeliger, P.~Vontobel, A.~Kavcic, and W.~Zeng,
  ``Simulation-based computation of information rates for channels with
  memory,'' \emph{{IEEE} Trans. Inf. Theory}, vol.~52, no.~8, pp. 3498--3508,
  Aug. 2006.

\bibitem{rusek12-01a}
F.~Rusek, A.~Lozano, and N.~Jindal, ``Mutual information of {IID} complex
  {Gaussian} signals on block {Rayleigh}-faded channels,'' \emph{{IEEE} Trans.
  Inf. Theory}, vol.~58, no.~1, pp. 331--340, Jan. 2012.

\end{thebibliography}
\end{document}